\documentclass[letter]{aa} 
\usepackage{graphicx}
\usepackage{txfonts}
\def\Mdot {$\dot M$\,}

\def\simlt{\lower.5ex\hbox{$\; \buildrel < \over \sim \;$}}
\begin{document}
\title{Signature of wide-spread clumping in B supergiant winds}
\subtitle{}
   \author{R.K. Prinja
           \inst{1} 
           \and
           D.L. Massa\inst{2}
}

   \institute{Department of Physics \& Astronomy, UCL, Gower Street, 
              London
              WC1E 6BT, UK\\
             \email{rkp@star.ucl.ac.uk}
        \and Space Telescope Science Institute, 3700 San Marino Drive,
               Baltimore, MD 21218, USA\\
	     \email{massa@stsci.edu}
	}

\date{Received July, 2010; accepted July, 2010}


\abstract
{}
{We seek to establish additional observational signatures
of the effects of clumping in OB star winds. The action of
clumping on strategic wind-formed spectral lines is
tested to steer the development of models for clumped
winds and thus improve the reliability of mass-loss determinations
for massive stars.}
{The Si{\sc iv} $\lambda\lambda$1400 resonance line doublets of
B0 to B5 supergiants are analysed using empirical line-synthesis
models. The focus is on decoding information on wind clumping
from measurements of ratios of the radial optical depths
($\tau_{\rm{rad}}(w)$)
of the red and blue components of the Si{\sc iv} doublet.
We exploit in particular the fact that the two doublet
components are decoupled and formed independently for targets with 
relatively low wind terminal velocities.}
{Line-synthesis analyses reveal that the mean ratio of
$\tau_{\rm{rad}}(w)$
of the blue to red Si{\sc iv} components are rarely close
to the canonical value of $\sim$ 2 (expected from atomic constants),
and spread instead over a range of values between $\sim$1 and 2.
These results are interpreted in terms of a photosphere
that is partially obscured by optically thick structures in the
outflowing gas.}
{The spectroscopic signatures established in this study
demonstrate the wide-spread existence of wind clumping
in B supergiants. The additional information in unsaturated doublet
profiles provides a means to quantify the porosity of the winds.}



\maketitle

\section{Introduction}
There are current serious uncertainties in the mass-loss and
thus energy feedback processes of massive OB stars. Over the
past decade evidence has accumulated from a broad collection of
results to strongly challenge the current model of hot star mass-loss
via stellar winds. Of particular impact on mass-loss determinations
is the presence of (small-scale) clumping
and (large-scale) structure, resulting in porous winds.
The
characteristics of the X-ray emission (e.g. Cohen, 2008),
excess flux at IR and mm wavelengths (Runacres {\&} Blomme 1996),
discordance between FUV (sensitive to density, $\rho$) and
H$\alpha$ (sensitive to $\rho^2$) measurements of mass-loss rates
(e.g. Massa et al. 2003; Bouret et al. 2005; Fullerton et al. 2006;
Puls et al. 2006) collectively suggest the winds are highly
clumped, with a consequent reduction in actual mass-loss rates.
Given that some of these studies question mass-loss rates of massive
stars at the order-of-magnitude level, there is the highest
urgency to robustly investigate this discordance since it has
potentially far reaching consequences for the evolution and
fate of massive stars (which is largely dependent on mass-loss)
and galactic chemical evolution (where mass-loss drives feedback
into the interstellar medium.)

Furthermore, recent theoretical predictions of the effects
of structure in porous winds (Oskinova et al. 2007; Sundqvist
et al. 2010), ionization by shock produced XUV line
radiation (Waldron {\&} Cassinelli 2010) and
X-rays (Krticka et al. 2009) have amplified the need to
verify via additional methods the compelling FUV studies cited
above, which were based on the P{\sc v} $\lambda\lambda$1118, 1128
doublet.

\section{The
Si{\sc iv} $\lambda\lambda$1400 resonance line doublet ratios}

In this Letter we report on additional, untapped information
on the nature of wind clumping that can be extracted from an
analysis of the doublet ratios of
Si{\sc iv} $\lambda\lambda$1393.76, 1402.77 in B supergiants.
The AGN community have previously
recognised the potential of ratios of doublet components as
a signature of clumping (e.g. Ganguly et al. 1999).
The reason is that for a source obscured by a {\it uniform
distribution} of gas, the ratio of optical depths determined
from the observed absorption lines would equal the ratio of
the oscillator strengths of the two doublet components.
However, in the case of a source that is partially obscured by
totally opaque clumps, the ratios of observed optical depth
will be 1, since they will depend only on the fraction
of the source which is uncovered.

\begin{figure*}
\includegraphics[scale=0.55]{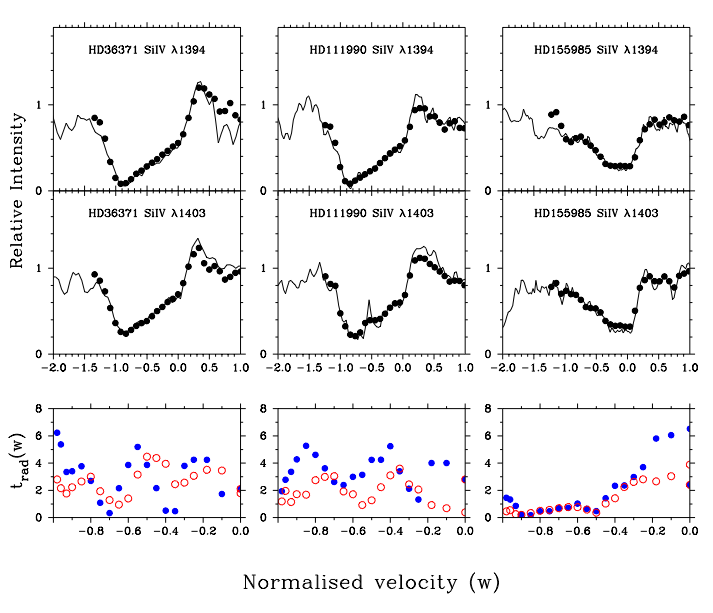}
\caption{Examples of SEI line synthesis fits to Si{\sc iv} 
$\lambda$1393.76 (upper panels)
and Si{\sc iv} $\lambda$1402.77 (middle panels) 
in our core B supergiant sample.
The lower-most panels show the corresponding radial optical
depths for the blue (filled circles) and red (open circles)
doublet components.}
\end{figure*}

We exploit this diagnostic here (in an extension to Massa et al. 2008):
In a homogeneous massive star wind, the ratio of the optical
depths of the widely spaced Si{\sc iv} $\lambda\lambda$1400
doublet would be the usual ratio of the oscillator strengths
of the two components (i.e. $\sim$ 2.01). However, in an OB
star wind composed of very optically thick clumps embedded in
a transparent medium, the radial optical depth of an absorption
line depends only on the geometric covering factor of the
clumps. Therefore in this case reducing the optical depths
of the clumps by a factor of 2 (as implied in the Si{\sc iv} doublets)
makes minimal difference in the absorption.
Thus in this clumped case the ratio of doublet optical depths
measured from the absorption lines would lie between $\sim$ 1 and 2. 

\begin{table*}[t]
\caption{Si{\sc iv} SEI model profile fitting parameters
for the core sample.}
\label{log1}
\begin{tabular}{llllllccc}
\hline
\hline
\multicolumn{1}{l}{Star}
&\multicolumn{1}{l}{Sp. type}
&\multicolumn{1}{l}{T$_{\rm eff}$ (kK)}
&\multicolumn{1}{l}{$v_\infty$}
&\multicolumn{1}{l}{$\beta$}
&\multicolumn{1}{l}{$v_{\rm turb}$}
&\multicolumn{1}{l}{$\tau_{\rm {rad}}(w)$(blue/red)}
&\multicolumn{1}{l}{$<$\Mdot$q_i$$>$(blue)}
&\multicolumn{1}{l}{$<$\Mdot$q_i$$>$(red)}
\\                                 
 & & & & & & &(10$^{-9}$ M$_\odot$ yr$^{-1}$) & (10$^{-9}$ M$_\odot$ 
yr$^{-1}$)
\\
\hline
\\

HD 13866    & B2 Ib-II:p  & 18.8 & 860  &   0.7  &  0.03 & 1.45 & 0.52 &  
0.70
\\
HD 14134    & B3 Ia & 16.5 & 495  &   0.5  &  0.10 & 1.37 & 0.83 &  1.39 
\\
HD 14143  &  B2 Ia  & 18.0  & 525  &   3.5  &  0.25 & 1.80 & 7.46 & 8.36 
\\
HD 36371    &  B4 Ib  & 16.5    &  465 &   0.8 &  0.17 & 1.10 & 0.38 &  
0.46 \\
HD 42087    &  B2.5 Ib  & 18.0 &  630 &    1.2 &  0.11 & 1.04 & 0.66 &  
1.24 \\
HD 43384    &  B3 Iab & 16.5 &  630 &    0.9 &  0.20 & 1.19 & 1.11 & 2.82 
\\
HD 47240  & B1 Ib  & 20.5 & 955  &   1.0  &  0.07 & 1.04 & 1.36 &  2.99 \\
HD 51309   & B3 Ib  & 16.5 &  730 &    0.5 &  0.15 & 1.55 & 0.94 &  1.81 
\\
HD 53138  & B3 Ia & 16.5 &  515 &    1.1 &  0.28 & 1.22 & 1.97 &  2.28 \\
HD 58350  & B5 Ia  & 15.0 &  380 &    0.8 &  0.25 & 2.04 & 0.90 & 1.25 \\
HD 58510    & B1 Ib-II & 20.5 & 915  &   0.9  &  0.19 & 1.79 & 1.37 & 1.62 
\\
HD 77581    & B0.5 Iae & 26.0  & 655  &   1.0  &  0.22 & 1.80 & 1.19 &  
1.72 \\
HD 79186    & B5 Ia & 15.0 & 435  &   0.8  &  0.23 & 1.73 & 0.86 &  1.09 
\\
HD 93827   & B1 Ib  & 20.5 & 590  &   0.5  &  0.16 & 1.64 & 0.10 & 0.15 \\
HD 111934    & B1.5 Ib & 18.5 & 935  &   3.1  &  0.15 & 1.46 & 4.62 &  
6.35 \\
HD 111973    & B3 Ia & 16.5 & 520  &   0.5  &  0.25 & 1.43 & 0.87 &  1.31 
\\
HD 111990   & B2 Ib & 18.0 & 715  &   3.3  &  0.12 & 1.97 & 2.14 &  2.50 
\\
HD 119608  & B1 Ib  & 20.5  & 925  &   0.6  &  0.25 & 1.14 & 2.05 & 3.69 
\\
HD 148379 & B2 Iap & 18.0 & 560  &   3.2  &  0.17 & 1.32 & 3.78 &  9.86 \\
HD 152236    & B0.5 Ia & 26.0 & 470  &   2.0  &  0.31 & 1.12 & 1.42 &  
2.03 \\
HD 152667    & B0 Ia & 27.5 & 800  &   2.5  &  0.28 & 1.78 & 4.44 &  4.81 
\\
HD 155985   & B0.7 Ib & 23.5 & 940  &   0.5  &  0.13 & 1.24 & 0.60 & 0.95 
\\   
HD 157246 & B1 Ib & 20.5 & 600  &   0.5  &  0.31 & 1.08 & 0.67 & 1.56 \\
HD 198478    & B2.5 Ia & 17.5 & 560  &   2.0  &  0.24 & 1.34 & 2.14 &  
4.16 \\
HD 225094    & B3 Ia & 16.5 & 485  &   0.5  &  0.21 & 1.75 & 0.81 & 0.78 
\\

\hline
\end{tabular}
\newline
$T_{\rm eff}$ from Searle et al. (2008). The $<$\Mdot$q_i$$>$ values
include a few cases where
$\tau_{\rm {rad}}(w)$ exceeds 5.0 in isolated velocity bins.
\end{table*}

\section{The B supergiant sample}
The dataset of Si{\sc iv} resonance line doublets analysed
in this study is derived from the B0 to B5 supergiants
discussed by Prinja et al. (2005), using high resolution
{\it International Ultraviolet Explorer} ($IUE$) satellite spectra.
In particular our {\it core sample} of targets are B supergiants that (i)
show well-developed but unsaturated Si{\sc iv} absorption troughs,
and (ii) have wind terminal velocities that are less than 0.5 of
the Si{\sc iv} doublet separation; i.e. stars with $v_\infty$
$\simlt$ 970 km s$^{-1}$. This is an important selection criterion
since it means that the two components of the doublet can be treated
as radiatively decoupled and can be analysed separately
i.e. each component as a singlet. The core sample of 25 stars
fitting our primary criteria above are listed in Table~1.

To extract physical information from the line profiles we
employed the 'Sobolev with Exact Integration' (SEI) code of
Lamers et al. (1987), but with the modifications described by
Massa et al. (2003), which permit the radial optical depth
($\tau_{\rm{rad}}(w)$) to be treated as 21 discrete,
independent and variable bins in velocity space. The bins are
then adjusted to match the observed profile by a non-linear
least squares procedure. The often complex photospheric spectrum
in the region of Si{\sc iv} $\lambda\lambda$1400 was included in
the SEI fitting by using the $IUE$ spectra of {\it dwarf}
B stars listed in Prinja et al. (2002). (The B dwarf spectra
do not exhibit Si{\sc iv} wind lines and were rotationally
broadened to match the target B supergiants.)

By model fitting the blue and red Si{\sc iv} components as singlets,
we are then able to extract two independent sets of
$\tau_{\rm{rad}}(w)$, which can be inspected for signatures
of wind clumping (as justified in Sect. 2). The SEI model fitting 
parameters,
including $v_\infty$, velocity law index ($\beta$) and turbulent
velocity parameter ($v_{\rm turb}$) are also listed in Table~1,
for the 25 B supergiants for which the Si{\sc iv} doublet components
can be considered as decoupled.

\begin{figure}
\includegraphics[scale=0.56]{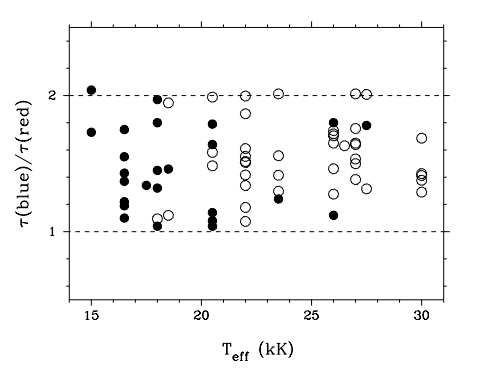}
\caption{{\it Filled circles} $-$ The mean optical depth ratio between 
0.2 and 0.9
$v_\infty$ for Si{\sc iv} $\lambda$1394/Si{\sc iv} $\lambda$1403
versus effective temperature, for our core sample of
25 stars, where $v_\infty$ is less than half the doublet
separation.
Only optical depths in the range 0.3 to 5.0 are considered, to avoid
poor quality and saturated cases, respectively. \newline
{\it Open circles} $-$ Plot of the derived oscillator strength ratios
versus effective temperature for additional B supergiants (Sect. 4.1),
where $v_\infty$ {\it exceeds} more than half the Si{\sc iv}
doublet separation. In this case, the ratio of
oscillator strengths was a free parameter in the fits.
}
\end{figure}

\section{The evidence for clumping}
Three representative examples of the Si{\sc iv} blue and red
components independently matched by SEI models are shown in
Fig. 1, covering the spectral type range B0.7 to B4.
The corresponding radial optical depths are also plotted for
each case. The ratio of the blue-to-red optical depths,
averaged between 0.2 to 0.9 $v_\infty$, is shown for the
full sample of 25 stars in Fig. 2 (filled circles; listed in
Table 1).
For reliability, we only
consider optical depths in the interval
0.3 $\le$ $\tau_{\rm rad}(w)$ $\le$ 5.0, to avoid weaker and
saturated lines.
We assign an error of $\pm$5{\%} in the $\tau_{\rm rad}(w)$ values
adopted in each velocity bin.
It is clear that in almost all cases the best fits for the decoupled 
doublet components yield optical depth ratios less than 2.01 (the 
expected value from atomic constants). The majority of
ratios lie between $\sim$ 1.0 and 1.5, and the overall mean
is $\sim$ 1.46 (s.d. $\sim$ 0.31). As argued in Sect. 2,
blue-to-red Si{\sc iv} doublet $\tau_{\rm rad}(w)$ ratios
significantly lower than $\sim$ 2 imply that the B supergiant
{\it winds are porous and contain very optically thick clumps.}
Note in Fig. 2 that the ratios do not depend on the
effective temperature of the relatively wide B0 to B5
spectral type range, and we therefore rule out any systematic
effects that would cause all of the derived ratios to differ
in the same manner. Our results clearly indicate that the
weaker (red) component of the Si{\sc iv} doublet is stronger
than expected from smooth wind models.

An alternative measure of the departure from the smooth wind
predictions is listed in Table 1, where the
product of mass-loss
rate $\times$ Si$^{3+}$ ion fraction (\Mdot$q_i$) is
provided for each doublet component, as determined from
the SEI fits (see formulation in Massa et al. 2003).
(The values in Table~1 are integrated over
0.2 $\le$ $v/v_\infty$ $\le$ 0.9).
In almost all cases the ratios of the blue to red                               
$<\dot{M}q>$  measurements lie between 0.5 and 1.0. These                       
results differ from the optical depth ratios shown in                           
Figure 2 because $\dot{M}q = const \; x^2 \;                       
w \; \frac{dw}{dx}\; \tau(w)/\lambda f$ (where $const$ contains                           
stellar and atomic data which are the same for both                             
components).  This relation has two consequences.  First,                       
the ratios of the $<\dot{M}q>$ scale as $(\lambda f)_{blue}/                    
(\lambda f)_{red} \simeq 2$ times the ratio of the mean                         
optical depths, which accounts for difference in range.                         
Second, the determinations of the mean values in Table 1
are weighted differently, which                           
accounts for the fact that mean                            
$\tau_{\rm{rad}}(w)$ ratio entry is not exactly 2 times                                
$<\dot{M}q>_{blue}/<\dot{M}q>_{red}$.

\subsection{Additional data sample}
The evidence for widespread wind clumping in B supergiants
presented here (Fig. 2) is based on the robust case of well separated,
decoupled Si{\sc iv} components, i.e. for stars that have
$v_\infty$ $\le$ 970 km s$^{-1}$. There are of course many
additional B supergiants with available UV data that have
$v_\infty$ in excess of $\sim$ 1000 km s$^{-1}$. As the terminal
velocity in these cases exceeds more than half the Si{\sc iv}
doublet separation, their doublet components cannot be matched
independently as 'singlets' (as above). We can nevertheless
extract some complementary information from these higher
$v_\infty$ stars.

From the original B supergiant $IUE$ sample of Prinja et al.
(2005), we selected an additional 41 B0 to B2 stars and fitted
the (now coupled) Si{\sc iv} doublets with the SEI models
(see Massa et al. 2008).
However, in these cases the ratio of the doublet oscillator
strengths was allowed to be a free parameter in the least squares
fits. The best fits to the sample of overlapping Si{\sc iv}
doublets yielded f-value ratios that also lie between
$\sim$ 1 and 2 in almost all cases. The results are also shown
in Fig. 2 (open circles) as a function of $T_{\rm eff}$.
For a fixed mass-loss rate, velocity law and ion fraction, the ratio of
blue-to-red oscillator strengths is equivalent to the
ratio of optical depths. Therefore the results shown in Fig. 2
are consistent with the original sample (Table 1 and Fig. 2
filled circles)
of low $v_\infty$, decoupled doublets.
Taken together, we conclude that the two datasets make a
compelling case for wide-spread clumping and porosity
in B supergiant winds.

\section{Discussion}
We have demonstrated that the Si{\sc iv} $\lambda\lambda$1400
doublets of B supergiants contain additional signatures of
clumping in the stellar winds. Line synthesis models of
25 B0 to B5 stars for which the Si{\sc iv} blue and red
components can be treated as decoupled ('singlets'), yield
blue-to-red optical depth ratios between 1 and 2, i.e.
mostly well below the value $\sim$ 2 expected from atomic
constants and a smooth wind. These results may be interpreted
as evidence for optically thick clumps in the wind, covering
only a fraction of the stellar source.
Additional results derived from B stars with overlapping
Si{\sc iv} doublets are consistent with this conclusion.

The optical depth ratios (Fig. 2) are not dependent
on the effective temperature. The large scatter observed in the ratio
for a given spectral type bin is also not correlated with the
projected stellar rotation velocity or mass-loss rate.
Aside from inclination
effects connected to large-scale structures such as co-rotating
interaction regions, the scatter in Fig. 2 may
reflect transient and episodic events in the wind.
It may be expected that for highly porous and
extremely clumped cases, the temporal effect due to the time-dependent
arrangement of clumps would be more pronounced.
As an illustration, we examined the $IUE$ time-series dataset of
HD~47240 (B1 Ib; Table 1) described by Prinja et al. (2002),
to test for temporal variations in the mean blue-to-red optical depth
ratios. Thirteen Si{\sc iv} spectra (SWP48835 to 48953),
spanning $\sim$ 16 days, were
matched by SEI models according to the methods for decoupled
doublets outlined in Sect. 3. The mean blue-to-red $\tau_{\rm{rad}}(w)$
ratio measured over 0.2 to 0.9 v$_\infty$ varies considerably during
the 16 day monitoring of HD~47240, with values between
$\sim$ 1.2 to 2.0, i.e. essentially the full range of scatter seen
in Fig. 2 at a given spectral type.
Consequently, we suspect that much of the scatter seen in Fig. 2
results from time-variable covering factors.
The spectroscopic changes modelled here in HD~47240 are empirically
due to time-variable velocity dependent structure in the UV absorption 
line profiles. Several previous time-series studies of variations
in the UV resonance lines of OB stars have interpreted this temporal
behaviour as due to the evolution and velocity migration of
large-scale structure in the winds, perhaps due to density
inhomogeneities or velocity plateaus. The substantial UV
line profile changes evident in OB stars are {\it not} diagnostics
of the effects of small-scale mirco-clumping. The inference here
then may be that the variable blue-to-red $\tau_{\rm rad}(w)$
ratios derived from the UV resonance lines of an individual OB star
reflect the porosity of a macro-clumped wind.
We now plan additional studies to derive information about the
spatial extent and temporal evolution of the obscuring structure,
which may be decoded from the suspected wavelength dependence
of observed changes in the mean $\tau_{\rm{rad}}(w)$ ratios.

\begin{figure}
\includegraphics[scale=0.38]{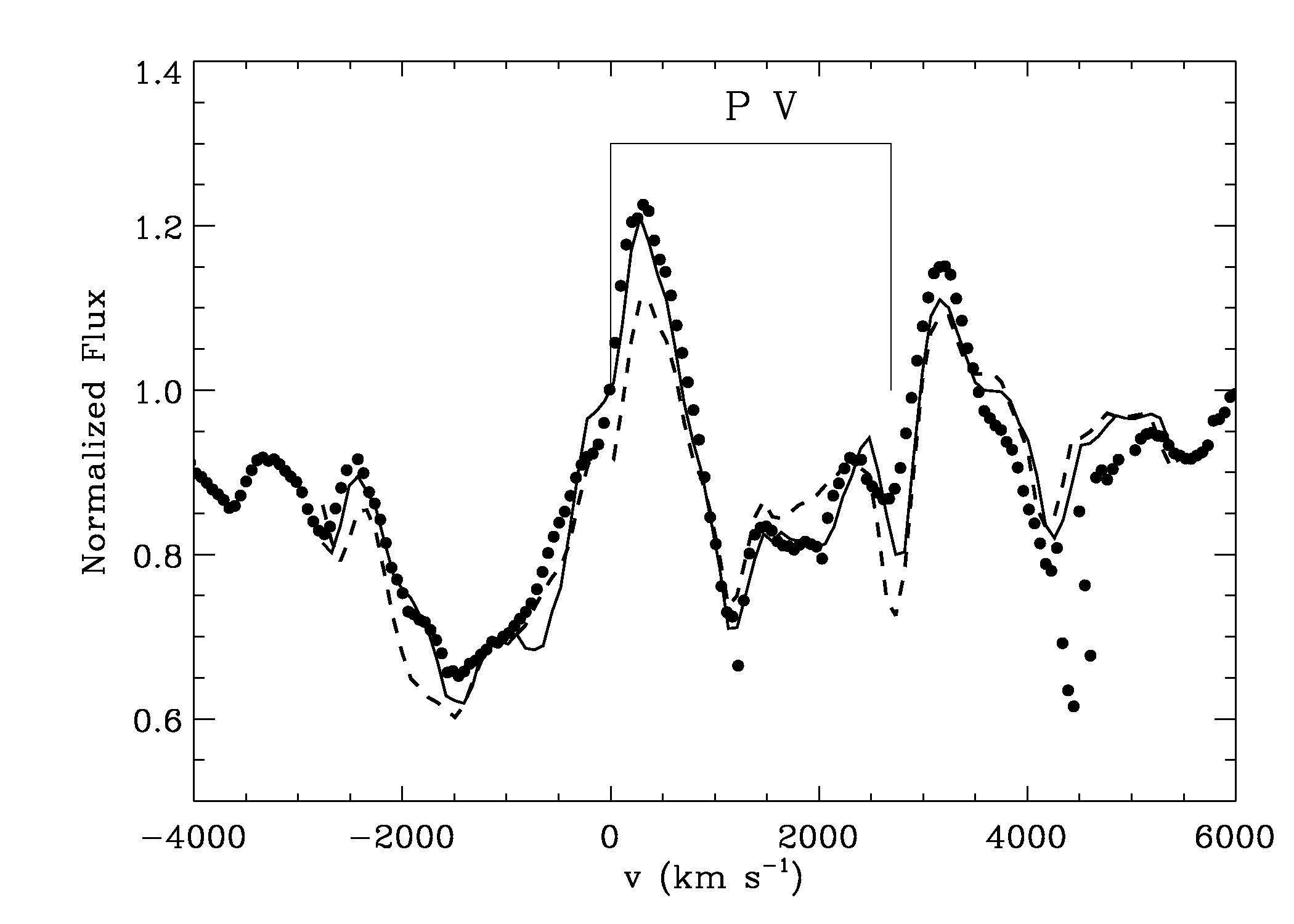}
\caption{Plot of the observed P{\sc v} $\lambda\lambda$1117, 1128
resonance line doublet in the O4 supergiant $\zeta$ Pup
(dotted line).
The dashed line fit is for the oscillator strength ratio
fixed by the atomic constants (= 2.02). The solid line is
the improved match obtained by allowing the f-ratio to
be a free parameter (=1.84 for the minimized residuals).
}
\end{figure}

To a large degree the current debate concerning clumping in
porous winds, and the consequent over-estimation of
clumping independent mass-loss rates, is founded on observational
and theoretical studies of O-type stars (see e.g. references in
Sect. 1). 
The line doublets diagnostics
that we have demonstrated here for B supergiants can to some extent
also be tested on O-type stars. Since the terminal velocities of O-type 
stars are mostly higher than for B supergiants, only the method
for overlapping doublets employed in Sect. 4.1 is feasible for
O stars, i.e. allowing the f-ratio of the doublet to vary from
the value expected from atomic constants. A demonstration of this
application is shown in Fig. 3 for the O4 I star, $\zeta$ Pup
from Massa et al. (2008).
We compare here matches to the FUV P{\sc v} $\lambda\lambda$1117, 1128
doublet P~Cygni profile for (i) the f-ratio fixed to the value
for the atomic constants (= 2.02), and (ii) allowing the
f-ratio to be a free parameter; a value of 1.84 is adopted for
the final least squares fit. This 10{\%} change in the f-ratio
undoubtedly improves the match, and is consistent with the evidence
for clumping in B supergiants reported in this Letter.
Finally, we anticipate that the clumping signature contained
in the optical depth ratios of doublet lines may also be exploited
in other low $v_\infty$ settings, such as the fast winds of
PN central stars and OB stars in the Magellanic Clouds.


\begin{acknowledgements}

DM acknowledges support from
NASA ADP contract NNH08CD06C to SGT, Inc.

\end{acknowledgements}

{}


\begin{thebibliography}{}

\bibitem {} Bouret, J.-C., Lanz, T., Hillier, D.J. 2005, A{\&}A, 438, 301

\bibitem {} Cohen, D.J. 2008, in Massive Stars as Cosmic Engines,
Proc. IAU Symposium, Vol. 250, p. 17-24, eds. F. Bresolin, P.A. Crowther, 
{\&} J. Puls,  Cambridge University Press

\bibitem {} Fullerton, A.W., Massa, D.L., Prinja, R.K. 2006, 
ApJ, 637, 1025

\bibitem {} Ganguly, R., Eracleous, M., Charlton, J.C.,
Chruchill, C.W. 1999, AJ, 117, 2594

\bibitem {} Krticka, J., Feldmeier, A., Oskinova, L.M.,
Kubat, J., Hamann, W. -R. 2009, A{\&}A, 508, 841

\bibitem {} Lamers, H.J.G.L.M., Cerruti-Sola, M., {\&} Perinotto,
M. 1987, ApJ, 314, 726

\bibitem {} Massa, D., Fullerton, A.W., Sonneborn, G., Hutchings,
J.B. 2003, ApJ, 586, 996

\bibitem {} Massa, D. L., Prinja, R. K., Fullerton, A. W. 2008,
in Clumping in Hot-Star Winds, ed. W.-R. Hamann, A. Feldmeier, {\&}
L. M. Oskinova, 147, Univ.-Verl.

\bibitem {} Oskinova, L.M., Hamann, W.-R., Feldmeier, A. 2007, A\&A,
476, 1331

\bibitem {} Prinja, R.K., Massa, D., Fullerton, A.W. 2002,
A{\&}A, 388, 587

\bibitem {} Prinja, R.K., Massa, D., Searle, S.C. 2005,
A{\&}A, 430, L41

\bibitem {} Puls, J., Markova, N., Scuderi, S.,
Stanghellini, C., Taranova, O., Burnley, A.W.,
Howarth, I.D. 2006, A{\&}A, 454, 625

\bibitem {} Runacres, M.C., Blomme, R. 1996, A{\&}A, 309, 544

\bibitem {} Searle, S.C., Prinja, R.K., Massa, D.,
Ryans, R. 2008, A{\&}A, 481, 777

\bibitem {} Sundqvist, J., Puls, J., Feldmeier, A. 2010,
A{\&}A, 510, 11

\bibitem {} Waldron, W.L., Cassinelli, J.P. 2010,
ApJ, 711, L30

\end{thebibliography}
\end{document}